\UseRawInputEncoding
\documentclass[
]{ceurart}

\usepackage{listings}
\usepackage{comment}
\begin{document}

\copyrightyear{2026}
\copyrightclause{Copyright for this paper by its authors.
  Use permitted under Creative Commons License Attribution 4.0
  International (CC BY 4.0).}

\conference{The 1st Late Interaction Workshop (LIR) @ECIR 2026}

\title{Spike Hijacking in Late-Interaction Retrieval}


\author[1]{Karthik Suresh}[%
orcid=0000-0002-0877-7063,
email=karsures@adobe.com,
]
\address[1]{Adobe, 345 Park Ave, San Jose, CA 95110, USA}

\author[2]{Tushar Vatsa}[%
orcid=0000-0001-7116-9338,
email=tvatsa@adobe.com,
]

\author[3]{Tracy King}[%
email=tking@adobe.com,
]

\author[3]{Asim Kadav}[%
email=akadav@adobe.com,
]

\author[3]{Michael Friedrich}[%
email=mifriedr@adobe.com,
]



\begin{abstract}
Late-interaction retrieval models rely on hard maximum similarity (MaxSim) to aggregate token-level similarities. Although effective, this winner-take-all pooling rule may structurally bias training dynamics. We provide a mechanistic study of gradient routing and robustness in MaxSim-based retrieval. In a controlled synthetic environment with in-batch contrastive training, we demonstrate that MaxSim induces significantly higher patch-level gradient concentration than smoother alternatives such as Top-k pooling and softmax aggregation. While sparse routing can improve early discrimination, it also increases sensitivity to document length: as the number of document patches grows, MaxSim degrades more sharply than mild smoothing variants.

\noindent We  corroborate these findings on a real-world multi-vector retrieval benchmark, where controlled document-length sweeps reveal similar brittleness under hard max pooling. Together, our results isolate pooling-induced gradient concentration as a structural property of late-interaction retrieval and highlight a sparsity--robustness tradeoff. These findings motivate principled alternatives to hard max pooling in multi-vector retrieval systems.

\end{abstract}

\maketitle

\section{Introduction}

Late-interaction retrieval models~\cite{Khattab2020ColBERT} have emerged as a powerful alternative to single-vector embedding methods~\cite{Karpukhin2020DPR,Reimers2019SentenceBERT}. Instead of compressing a query and document into single representations, late-interaction approaches retain token or patch-level embeddings and compute relevance by aggregating fine-grained similarity scores. A widely adopted aggregation rule in such models is hard maximum similarity (MaxSim), where each query token contributes the maximum similarity it achieves over document patches. This winner-take-all formulation has been shown to be both effective and computationally attractive in multi-vector retrieval systems~\cite{Khattab2020ColBERT,Santhanam2022ColBERTv2,Santhanam2022PLAID}.

Despite its empirical success, the structural implications of MaxSim remain underexplored. MaxSim introduces a non-smooth routing mechanism: for each query token, only the highest-scoring document patch contributes to the final score. Under contrastive training with in-batch negatives~\cite{vandenOord2018CPC}, this induces a highly selective gradient pathway. While sparsity can enhance discrimination, it can concentrate learning signal on a small subset of document patches,  affecting robustness and generalization.

We provide a mechanistic analysis of pooling in late-interaction retrieval, answering  questions:
\begin{enumerate}
    \item How does MaxSim based pooling shape gradient routing under contrastive learning?
    \item Does this structural property affect robustness, particularly as document length increases?
\end{enumerate}

To isolate the effect of pooling, we construct a controlled synthetic retrieval environment in which queries and documents are generated from a shared concept dictionary. Within this setting, we train a small encoder using in-batch InfoNCE~\cite{vandenOord2018CPC} and compare three aggregation strategies: MaxSim, Top-$k$ averaging, and temperature-controlled softmax pooling. This controlled setup allows us to measure patch-level gradient concentration and retrieval behavior as document length varies.

Our synthetic experiments reveal three findings. First, MaxSim induces significantly higher patch-level gradient concentration than smoother alternatives. Second, although MaxSim achieves competitive retrieval quality in moderate-length documents, it exhibits greater sensitivity to increasing document length. Third, mild smoothing (e.g., Top-$k$ pooling) preserves retrieval performance while reducing gradient concentration and improving length robustness. These results expose a sparsity--robustness tradeoff inherent in pooling design.
To validate that this  is not confined to the synthetic regime, we  conduct controlled document-length sweeps in a real-world multi-vector retrieval benchmark. The empirical trends mirror those observed in the synthetic environment, providing corroborating evidence that hard max pooling induces increased brittleness with respect to document length.

\noindent Our contributions are as follows:
\begin{itemize}
    \item A mechanistic study of gradient routing under hard max pooling in late-interaction retrieval.
    \item Introduction of patch-level gradient concentration metrics to quantify pooling-induced sparsity.
    \item Demonstration, in a controlled synthetic setting, that MaxSim leads to increased sensitivity to document length compared to mild smoothing alternatives.
    \item Empirical validation on a real-world benchmark showing that the sparsity-smoothness tradeoff governs pooling robustness across both adversarial and non-adversarial document-length increases.
\end{itemize}

\section{Background and Problem Formulation}

\subsection{Late-Interaction Retrieval}

Let a query $q$ consist of $T$ token embeddings
\[
q = \{ \mathbf{q}_1, \dots, \mathbf{q}_T \}, \quad \mathbf{q}_i \in \mathbb{R}^d,
\]
and let a document $d$ consist of $M$ patch embeddings
\[
d = \{ \mathbf{d}_1, \dots, \mathbf{d}_M \}, \quad \mathbf{d}_j \in \mathbb{R}^d.
\]

\noindent Late-interaction retrieval models~\cite{Khattab2020ColBERT} compute a similarity matrix
\[
S_{ij} = \mathbf{q}_i^\top \mathbf{d}_j,
\]
and aggregate these fine-grained similarities into a scalar relevance score
\[
\text{score}(q,d) = \sum_{i=1}^{T} \phi\big( S_{i1}, \dots, S_{iM} \big),
\]
where $\phi(\cdot)$ is a pooling operator applied independently for each query token.

\subsection{Pooling Operators}

We consider three commonly used pooling strategies.

\paragraph{Hard Max (MaxSim).} Each query token contributes only its highest similarity over document patches~\cite{Khattab2020ColBERT}.
\[
\phi_{\max}(S_{i1}, \dots, S_{iM}) = \max_{1 \le j \le M} S_{ij}.
\]

\paragraph{Top-$k$ Averaging.}  $\text{Top-}k(i)$ denotes the indices of the $k$ largest values in $\{S_{ij}\}_{j=1}^M$.
\[
\phi_{k}(S_{i1}, \dots, S_{iM}) = 
\frac{1}{k} \sum_{j \in \text{Top-}k(i)} S_{ij},
\]

\paragraph{Softmax Pooling.}  $\tau > 0$ controls smoothness. As $\tau \to 0$, softmax pooling approaches hard max.
\[
\phi_{\tau}(S_{i1}, \dots, S_{iM}) =
\sum_{j=1}^{M} w_{ij} S_{ij}, 
\quad
w_{ij} = \frac{\exp(S_{ij} / \tau)}{\sum_{m=1}^{M} \exp(S_{im} / \tau)},
\]

\subsection{Contrastive Training Objective}

We train models using in-batch InfoNCE~\cite{vandenOord2018CPC}. For a batch of $B$ query-document pairs $\{(q_i, d_i^+)\}_{i=1}^B$, the score matrix is defined as
\[
\mathbf{Z}_{ij} = \text{score}(q_i, d_j^+).
\]

The objective is
\[
\mathcal{L} = 
\frac{1}{B} \sum_{i=1}^{B}
- \log 
\frac{\exp(\mathbf{Z}_{ii})}
{\sum_{j=1}^{B} \exp(\mathbf{Z}_{ij})}.
\]

\noindent Each query is trained to assign higher score to its positive document relative to in-batch negatives.

\subsection{Patch-Level Gradient Concentration}

To quantify pooling-induced sparsity, we analyze how gradient mass is distributed across document patches.
For a fixed query-document pair $(q,d)$ and loss $\mathcal{L}$, let
\[
g_j = \left\| \frac{\partial \mathcal{L}}{\partial \mathbf{d}_j} \right\|
\]
denote the gradient norm for patch $j$.

We define patch-level gradient concentration using the Gini coefficient~\cite{Gini1912} over $\{g_j\}_{j=1}^{M}$:
\[
\text{Gini}(g) =
1 - \frac{2}{M \sum_{j=1}^{M} g_j}
\sum_{j=1}^{M}
\left( \sum_{m=1}^{j} g_{(m)} \right),
\]
where $g_{(m)}$ denotes the sorted gradient norms in ascending order.
A higher Gini value indicates that gradient mass is concentrated on fewer patches. This metric directly compares how different pooling operators route learning signal during contrastive training.

\subsection{Problem Statement}

Our goal is to understand how the choice of pooling operator $\phi$ affects:
(1) Patch-level gradient concentration under contrastive training, and
(2) Retrieval robustness as document length $M$ increases.
By isolating pooling within a controlled setting, we aim to characterize the structural tradeoffs between selectivity and smoothness in late-interaction retrieval.

\section{Synthetic Experimental Setup}

To isolate the structural effect of pooling, we construct a controlled synthetic retrieval environment. The design allows precise control over query--document alignment, document length, and negative hardness while keeping the training objective fixed. 

\paragraph{{Concept Dictionary and Embedding Space}} We generate a fixed dictionary of $C$ concept vectors 
\[
\{ \mathbf{c}_1, \dots, \mathbf{c}_C \} \subset \mathbb{R}^d,
\]
sampled from a standard Gaussian distribution and $\ell_2$-normalized.
Each query consists of $T$ token embeddings. A query is generated by sampling $T$ concept indices uniformly at random and adding Gaussian noise:
\[
\mathbf{q}_i = \text{normalize}(\mathbf{c}_{k_i} + \epsilon), 
\quad \epsilon \sim \mathcal{N}(0, \sigma_q^2 \mathbf{I}).
\]

\paragraph{Positive Documents} A positive document of length $M$ consists of $M$ patch embeddings. To control task difficulty, only $K \le T$ patches correspond to query concepts (distributed support), while the remaining patches are sampled from random concepts:
\[
\mathbf{d}_j^+ =
\begin{cases}
\text{normalize}(\mathbf{c}_{k_i} + \epsilon), & \text{if } j \le K, \\
\text{normalize}(\mathbf{c}_{r} + \epsilon), & \text{otherwise},
\end{cases}
\]
where $\epsilon \sim \mathcal{N}(0, \sigma_d^2 \mathbf{I})$ and $r$ is a random concept index.
This design ensures that positive relevance is distributed across multiple patches rather than concentrated in a single location.

\paragraph{Hard Negatives} To simulate realistic confusers, we construct hard negative documents that:
   Share a small number of query concepts (partial overlap) and
    contain a ``spike'' patch formed by averaging several query concepts.
 The spike patch increases similarity across multiple query tokens and creates a potential winner-take-all scenario under hard max pooling. The remaining patches are sampled from random concepts. 
While synthetic by construction, spike patches model a practically relevant scenario: in retrieval-augmented generation, adversarially crafted or SEO-optimized passages may aggregate multiple query-relevant terms into a short span, creating a similar winner-take-all vulnerability. Our setup isolates this mechanism as a controlled stress test rather than a claim about attack frequency.

\paragraph{Training Protocol}
We train a small three-layer feed-forward encoder $f_\theta$ that maps raw embeddings to normalized representations:
\[
\tilde{\mathbf{q}}_i = f_\theta(\mathbf{q}_i),
\quad
\tilde{\mathbf{d}}_j = f_\theta(\mathbf{d}_j).
\]

Training uses in-batch InfoNCE~\cite{vandenOord2018CPC} with batch size $B=64$, where each query competes against $B-1$ in-batch negatives. The dataset is fixed across epochs to ensure learnability.
Unless otherwise stated, we train for 100 epochs with Adam~\cite{Kingma2015Adam} and evaluate every 5 epochs.

\paragraph{Evaluation Metrics}
We report:

\begin{itemize}
    \item \textbf{Recall@$k$}: fraction of queries whose positive document is ranked in the top-$k$.
    \item \textbf{Patch-level Gradient Gini}: inequality of gradient mass across document patches (Section~2).
\end{itemize}

\paragraph{Document Length Sweep}
To study robustness, we vary the number of document patches $M \in \{32, 64, 128, 256\}$ while keeping:
     Query tokens fixed;
     Concept assignments fixed;
     Training procedure unchanged.
Only the number of document patches changes. This controlled sweep isolates the effect of document length on retrieval quality under different pooling operators.

\paragraph{Compared Pooling Variants}
We compare three variants. This setup enables controlled comparison of gradient routing behavior and retrieval robustness across pooling strategies.

\begin{itemize}
    \item MaxSim,
    \item Top-$k$ averaging (with fixed $k$),
    \item Softmax pooling with temperature $\tau = 1.0$.
\end{itemize}

\section{Results}

\begin{figure*}[htb]
    \centering
    \includegraphics[width=\textwidth]{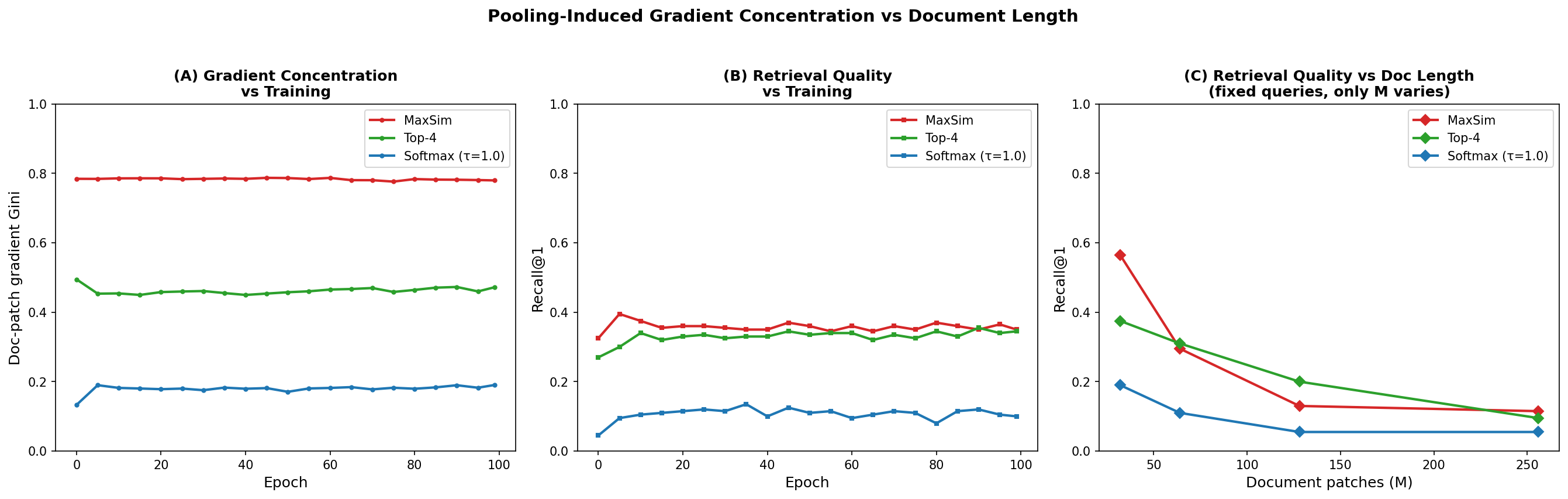}
    \caption{
      \noindent
    Synthetic training dynamics and document-length sweep.
    (A) Patch-level gradient concentration (Gini) vs. training.
    (B) Retrieval quality (Recall@1) vs. training.
    (C) Retrieval quality vs. document length $M$ under fixed queries.
    }
    \label{fig:synthetic}
\end{figure*}

We evaluate how pooling choice affects (i) gradient routing during training,
(ii) retrieval performance, and (iii) robustness to document length.
All synthetic results are summarized in Figure~\ref{fig:synthetic}.

\subsection{Synthetic Analysis}

\paragraph{Experimental Setup.}
We generate a fixed synthetic dataset of queries and positive documents built
from $C = 100$ latent concepts in $\mathbb{R}^d$ ($d = 16$). Each query contains
$T = 16$ concept tokens sampled with Gaussian noise $\sigma_q = 0.1$, and each
positive document contains $M$ patches with noise $\sigma_d = 0.1$, of which a
subset correspond to query concepts and the remainder are distractors.
Training is performed with in-batch negatives (batch size $B = 64$) under three
pooling strategies: hard MaxSim, Top-$k$ averaging ($k = 4$), and softmax
aggregation ($\tau = 1.0$).

To study robustness under adversarial length expansion, we inject
$K \in \{0, 25, 50, 100, 200, 400\}$ token-targeted hard-negative patches into
\emph{non-gold} documents only. We measure (i) patch-level gradient concentration
using the Gini coefficient and (ii) retrieval quality using Recall@1.

\paragraph{Gradient Concentration During Training.}

Figure~\ref{fig:synthetic}A shows the patch-level gradient Gini coefficient over training epochs.
MaxSim produces substantially higher gradient concentration than the alternatives. Throughout training, MaxSim maintains a Gini coefficient near $0.78$, indicating that gradient mass is concentrated on a small subset of document patches. Top-$4$ pooling yields moderate concentration ($\sim 0.45$), while softmax pooling distributes gradients far more uniformly ($\sim 0.18$).
Importantly, this separation emerges early and remains stable across epochs, suggesting that gradient concentration is a structural property of the pooling operator, not a transient optimization artifact.

\paragraph{Retrieval Performance During Training.}

Figure~\ref{fig:synthetic}B reports Recall@1 over training.
MaxSim and Top-$4$ pooling achieve comparable retrieval performance ($\sim 0.33$--$0.38$ Recall@1), while softmax pooling performs substantially worse. Thus, sparse routing appears beneficial for discrimination, but extreme sparsity (hard max) is not uniquely necessary for strong retrieval performance.

\paragraph{Robustness to Document Length.}

To evaluate brittleness, we conduct a controlled document-length sweep. Queries and concept assignments are fixed, while only the number of document patches $M$ varies.
Figure~\ref{fig:synthetic}C shows Recall@1 as a function of $M$.
All pooling strategies degrade as document length increases, reflecting increased opportunities for spurious matches. However, MaxSim exhibits sharper degradation compared to Top-$4$ pooling. While MaxSim performs strongly at small $M$, its performance drops more rapidly as $M$ grows. Top-$4$ pooling degrades more smoothly and outperforms MaxSim at larger document lengths.

\paragraph{Summary of Synthetic Findings.}

The synthetic results reveal a consistent pattern:

\begin{itemize}
    \item Hard MaxSim induces the highest gradient concentration.
    \item Moderate sparsity (Top-$k$ pooling) achieves similar retrieval quality with lower concentration.
    \item Higher concentration correlates with increased sensitivity to document length.
\end{itemize}

Together, these results isolate pooling-induced gradient routing as a key structural factor influencing robustness in late-interaction retrieval.
These results do not imply that lower concentration is uniformly desirable. In our synthetic setting, softmax pooling distributes gradients most evenly but yields the lowest retrieval quality
(Figure~\ref{fig:synthetic}B), suggesting that some degree of selective routing is
necessary for discrimination. The practical tradeoff is
regime-dependent: MaxSim excels on short, well-targeted documents,
while moderate smoothing (e.g., Top-$k$) offers a better
sparsity--robustness balance when document length is variable
or adversarial content is a concern.

\subsection{Real-World Corroboration on ColQwen2.5 + ViDoRe}
\label{sec:realworld}
\label{sec:realworld}

We validate our synthetic findings using ColQwen2.5~\cite{Faysse2024ColQwen2} on the ViDoRe biomedical benchmark~\cite{Faysse2025ColPali,Mace2025ViDoRev2}. Our goal is to test whether pooling-induced gradient concentration and spike hijacking manifest in practice.

Without retraining, we measure doc-patch gradient concentration (Gini) across 160 queries.
MaxSim produces the highest concentration (0.983), followed by Top-5 (0.951) and Softmax (0.883),
with paired $t$-tests confirming significant differences ($p < 10^{-80}$). This ordering matches
the synthetic results, indicating that concentration is structurally induced by pooling rather
than an artifact of the training regime.

To test robustness, we inject $K \in \{0, 25, 50, 100, 200, 400\}$ token-targeted hard-negative
patches into non-gold documents. Even at moderate injection levels ($K = 100$), MaxSim and Top-4
retain only 27.8\% and 29.3\% of baseline recall respectively, while Softmax retains 67.6\%.
Spike hijack rates reach approximately 0.70 for both MaxSim and Top-$k$, meaning most query
tokens select injected distractors. Notably, increasing $k$ (Top-1, Top-4, Top-16) does not
restore robustness: all spike-based operators retain only $\approx$28--30\% recall, because even
at higher $k$ the injected patches dominate the top-$k$ set and poison the average.

A Gaussian control experiment confirms this degradation is semantic rather than length-driven:
replacing hard negatives with random Gaussian patches preserves recall ($\approx$97\% retained,
hijack $< 0.15$). We note that these experiments swap pooling operators at inference time on a
MaxSim-trained model; end-to-end retraining with each pooling objective may further amplify
the benefit of smoother operators and is an important direction for future work.

Figure~\ref{fig:hijack_heatmap} illustrates this mechanism at the token level: under hard-negative
injection, $\sim$83\% of token-wise argmax selections shift to injected
patches, whereas Gaussian injection produces minimal routing disruption.

\paragraph{Summary.}
Real-world experiments confirm that (i) gradient concentration is structural, (ii) semantic distractors induce spike hijacking and recall collapse, and (iii) larger $k$ does not mitigate brittleness, closely mirroring synthetic findings.

\break
\bibliography{sample-ceur}

\break
\appendix

\section{Mechanistic Visualization of Spike Hijacking}
\label{app:hijack}

To complement aggregate retrieval metrics, we provide a qualitative 
token--patch similarity visualization illustrating the spike hijacking 
mechanism observed in real-world experiments.

\begin{figure*}[t]
\centering
\includegraphics[width=\textwidth]{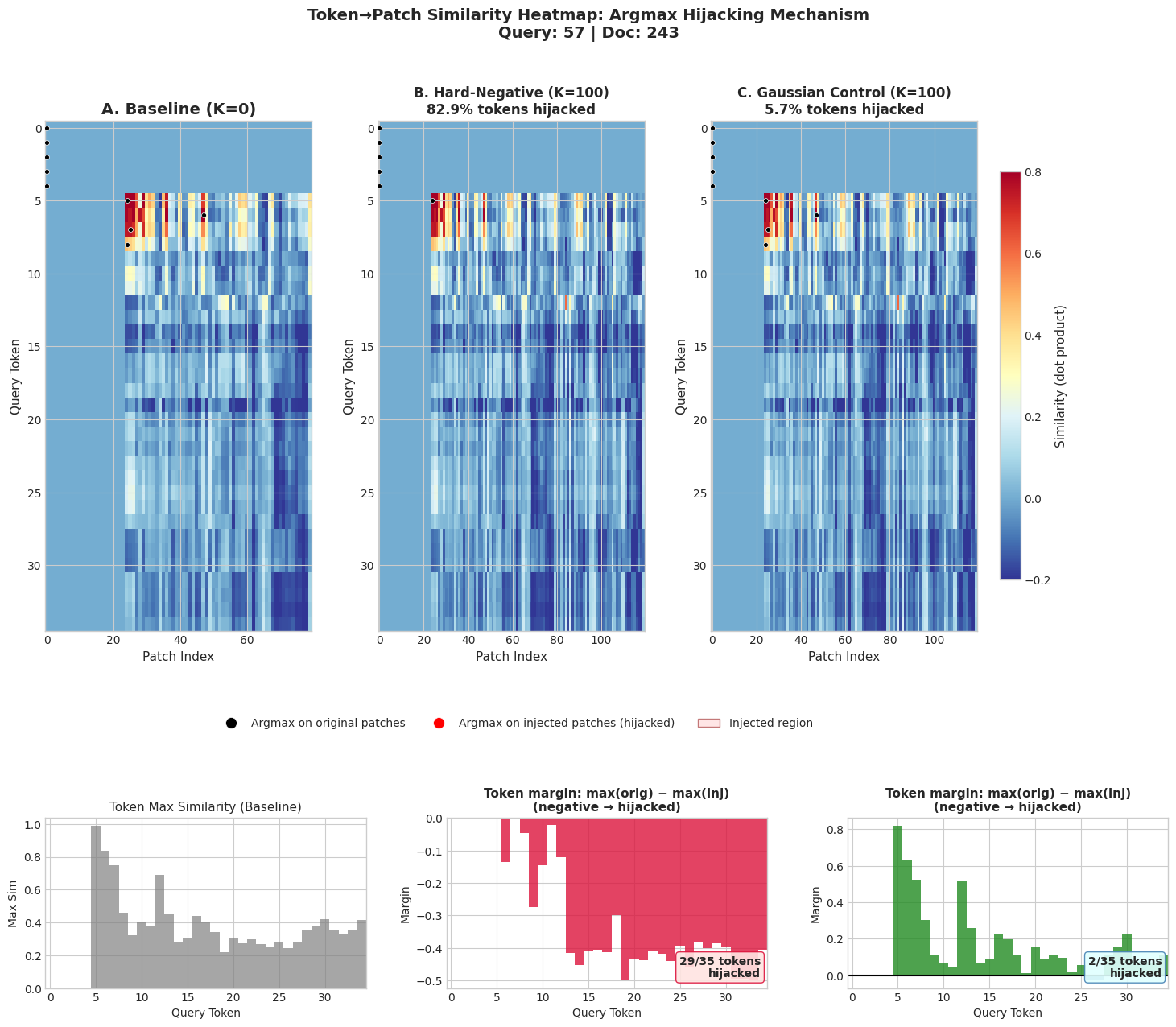}
\caption{Token--patch similarity heatmaps for a representative ColQwen2.5 example from ViDoRe biomedical retrieval. \textbf{Left (Baseline, $K=0$):} Token maxima align with semantically relevant document patches. \textbf{Middle (Hard-negative injection, $K=100$):} Injected distractor patches redirect the majority of token-wise argmax selections into the injected region, resulting in $\sim$83\% token hijacking. \textbf{Right (Gaussian control, $K=100$):} Random noise produces minimal routing changes and low hijack rates. Black dots denote argmax positions on original patches; red dots denote argmax positions on injected distractor patches. The injected region is highlighted.}
\label{fig:hijack_heatmap}
\end{figure*}

This visualization provides direct mechanistic evidence for the phenomenon 
quantified in Section~\ref{sec:realworld}: spike-based pooling routes token 
attention to high-similarity distractor patches, causing retrieval degradation, 
while soft aggregation avoids catastrophic token rerouting.

\end{document}